# Arsenal of Hardware Prefetchers

Dishank Yadav [*], Chaitanya Paikara [+],
[*]Indian Institute of Technology Kharagpur
[+]University of Washington

## ABSTRACT

Hardware prefetching is one of the latency tolerance optimization techniques that tolerate costly DRAM accesses. Though hardware prefetching is one of the fundamental mechanisms prevalent on most of the commercial machines, there is no prefetching technique that works well across all the access patterns and different types of workloads. Through this paper, we propose Arsenal, a prefetching framework which allows the advantages provided by different data prefetchers to be combined, by dynamically selecting the best-suited prefetcher for the current workload. Thus effectively improving the versatility of the prefetching system. It bases on the classic Sandbox prefetcher that dynamically adapts and utilizes multiple offsets for sequential prefetchers. We take it to the next step by switching between prefetchers like Multi look Ahead Offset Prefetching and Timing SKID Prefetcher on the run. Arsenal utilizes a space-efficient pooling filter, Bloom filters, that keeps track of useful prefetches of each of these component prefetchers and thus helps to maintain a score for each of the component prefetchers. This approach is shown to provide better speedup than anyone prefetcher alone. Arsenal provides a performance improvement of 44.29% on the single-core mixes and 19.5% for some of the selected 25 representative multi-core mixes.

## 1. INTRODUCTION

Most modern prefetchers are designed with a particular scenario in mind and thus give better performance only when the cache access pattern matches that scenario. In this work we present ARSENAL, a data prefetching framework that dynamically selects the best-suited prefetcher among its components for the current workload and deploys it, which ensures the highest possible speedup irrespective of the cache access pattern type. As proof of concept, we present two cases, One at L1D cache level and another at L2D.

## 1.1 Literature survey and Motivation

To understand the effectiveness of state-of-the-art prefetchers on a common scale framework, we analysed the performance of various L1 Cache centric prefetchers like TSKID[1], MLOP [2], Bingo[3] , pangloss[4] as well as L2 cache centric prefetchers like SPP[5], VLDP[6], Best-offset[7] using trace-based simulator, Champsim. We used traces for SPEC CPU 2017 to compare their performance. In the case of L1D centric prefetchers T-SKID comes up as a clear winner for the single-core mixes, in terms of overall performance; however, there are workloads in which TSKID underperforms compared to other prefetchers. For example, in cases of GCC and fotonik3d traces, MLOP provides greater speedup than SKID. Detailed analysis across the benchmark, as shown in Figure1, reveals that much higher a speedup if we pick the best performing prefetcher for each workload. A similar observation was made for L2D cache centric prefetchers, as summarised in Figure2. This lead to the inception of the idea to recognize the type of workload dynamically and deploy the suitable prefetcher on the run. For such a framework to provide the maximum benefit with a minimum number of component prefetchers (thus minimum overhead), the components chosen have to orthogonal i.e., give good performance for complementary sets of workloads. The analysis leads to the conclusion that SKID[1] and MLOP[2] formed such a pair among the L1d centric prefetchers, so these were chosen for the first test case targeting L1d cache. Among L2d centric prefetchers, SPP[5] and IP-stride were chosen.we have explored all the conventional and state-of-the-art prefetchers, and prefetchers that appear in Data Prefetching Championship 1 [8, 9, 10, 11, 12, 13, 14, 15], Data Prefetching Championship 2 [16, 17, 18, 19, 20, 6, 5, 7] , and Data Prefetching Championship 3 [21, 22, 23, 1, 4, 3, 2] to find this best combination. In this article, we present these two cases as proof of concept for the arsenal framework.

## 2. IMPLEMENTATION

In this section, we provide implementation details of the arsenal framework for both test cases. Portions with no distinction for the test case 1 and test case 2 are common for both.

## 2.1 Component Prefetchers

Here we introduce the different component prefetchers that we have analyzed and eventually used as a proof of concept for our Arsenal prefetching framework.

### 2.1.1 Test Case 1

**Timing SKID** [1] T-SKID prefetcher utilizes the repetitive access patterns spread over a larger instruction window, which the conventional prefetchers fail to recognize because of a short instruction window. Cache misses, even if predicted and prefetched successfully, maybe evicted before being accessed because of intermediary thrashing. The T-SKID learns these access patterns and effectively controls the prefetch tim-ing based on a PC, which has a strong correlation of memory access patterns even indifferent address zones.

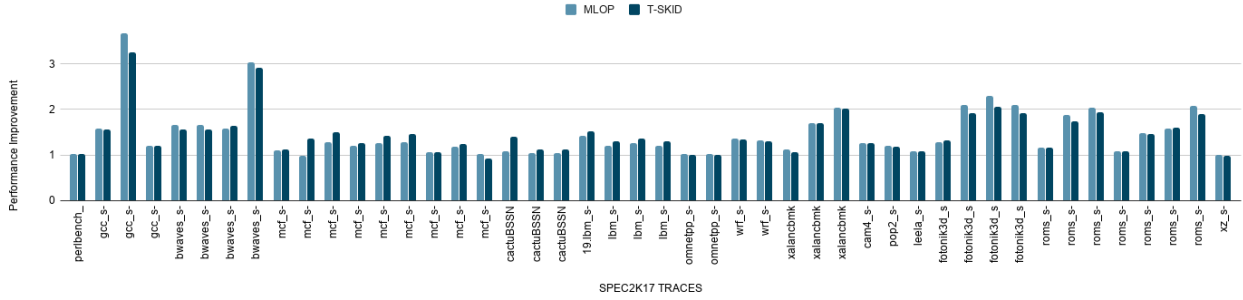

Figure 1: Normalized performance with different prefetchers for test case 1.

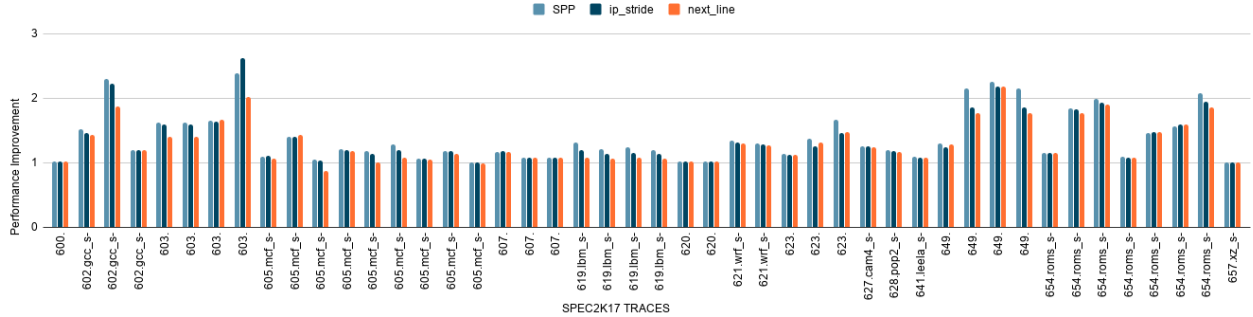

Figure 2: Normalized performance with different prefetchers for test case 2.

**Multi-Lookahead Offset Prefetcher**[2] Evaluates different prefetching offsets on the two metrics of timeliness and miss coverages, as many of the conventional offset prefetchers either neglect timeliness or sacrifice miss coverage while selecting the optimum offset for prefetch. The state of the art offset prefetchers generally lose on cache miss coverage because of their reliance on a single best offset, which generates most timely prefetch requests, however, instead of such a binary classification, MLOP considers multiple lookaheads for every prefetch offset and scores them individually. It then selects one offset for each lookahead level and thus allows prefetcher to issue enough requests while still considering the timeliness of these prefetch requests.

### 2.1.2 Test Case 2

**IP-stride** is a stride prefetcher that can handle stride patterns based on instruction pointer. It maintains a table of previous addresses accessed by a list of instruction pointers. When the same instruction is executed again, a stride is calculated in the address accessed and a prefetch request is made based on it. Replacement of stored IPs is based on LRU algorithm.

**Signature Path Prefetcher** (SPP) [5] stores the stride patterns in a compressed form in the signature table (ST). Each entry in the ST is used to index into the pattern table (PT), which is used to predict the next stride and also contains the confidence for the current prefetch. The signature is then updated with the latest stride and is used to recursively lookup the PT to predict more strides. This goes on until the confidence, which is multiplied with the last prefetch confidence goes below a certain threshold. The GHR stores prefetch requests that cross page boundaries so that prefetching can take place across pages.

**Next line** is one of the simplest prefetchers which prefetches the next cache line on each cache miss or prefetch hit. Here we used a modified version, which varies its aggressiveness or the number of cache lines prefetched based on its score.

## 2.2 The Arsenal Framework

### 2.2.1 Gathering performance parameters

Arsenal is motivated by the basic sandbox prefetcher[24], which searches and selects the best offset for an application. With Arsenal, we try to select the best prefetcher among the available components. The Arsenal framework is trained with prefetch activation events(PAE), i.e., cache Misses and cache prefetch hits. The framework works in two phases: (i) a continuous evaluation phase and (ii) a selection phase. The selection phase is triggered when the evaluation count (number of prefetcher calls) of all the component prefetchers crosses a threshold, which is considered after careful examination. At the end of every selection phase, the best-suited prefetcher is selected using the parameters gathered during the evaluation phase (in some cases, none might get selected). At each PAE, all the prefetchers are triggered by the Arsenal framework. Cache lines prefetched by the prefetchers are stored in their respective boom filters [25] without passing them along to the prefetch queue. Only the prefetch requests of the prefetcher that is selected during the last selection phase, are passed to the prefetch queue i.e.actually prefetched. Also, the evaluation counter of each of the prefetchers is incremented by one, and the prefetch count is incremented by the number of prefetch requests. At every miss and prefetch hit, the cache line address corresponding to the demanded address is compared to the contents of each of the Bloom filters. If any of the Bloom filters produce a match, then the corresponding prefetcher score is incremented by SCORE-INC; otherwise, the score is decremented by SCORE-DEC. A switch from the evaluation to the selection phase happens when all of the evaluation counters exceed EVAL-CNT.

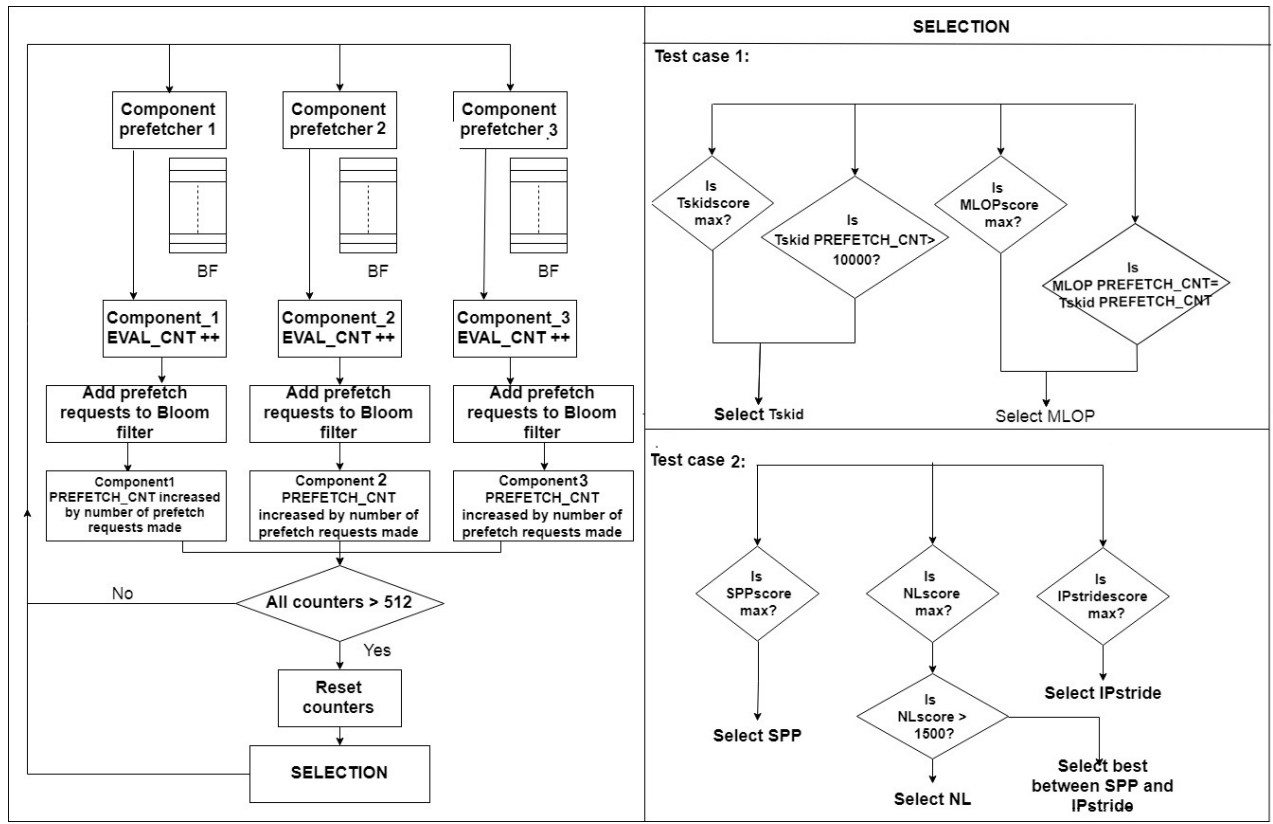

Figure 3: The Arsenal framework: evaluation phase and selection.

## 2.3 Prefetcher Selection

### 2.3.1 Test case 1

As T-SKID and MLOP are intelligent prefetchers that adjust their own aggressiveness based on a feedback mechanism number of prefetches attempted by these are also considered in addition to prefetchers' scores. Specifically, when a wrong (less favorable) prefetcher is selected, it's score might get inflated, leading to a faulty cycle where the wrong prefetcher will keep getting selected. The number of prefetches attempted can be used to correct this. If T-skid score is higher or if T-skid attempts more prefetches than TSKID_SELECTION_ATTEMPT T-skid is selected. If MLOP score is higher or if MLOP prefetch attempts are equal to Tskid, MLOP is selected.

### 2.3.2 Test case 2

If SPP or IP-stride have the maximum score and it is greater than the MIN-SCORE threshold, the prefetcher is selected. If the score of the next line is the maximum among the three and it is greater than a threshold called NEXT-LINE-MIN-SCORE, then the next line is selected. If the score of next-line is the maximum of the three, but it is not higher than NEXT-LINE-MIN-SCORE, the one among SPP andIP-stride, which has a higher score (and also greater thanMIN-SCORE) is selected. If none of the scores cross their respective threshold, then no prefetcher is selected.

Figure3 illustrates the evaluation and the selection phase of our Arsenal framework.

**Thresholds of interest:** These are some of the thresholds used by Arsenal framework in the presented work. score increment on a prefetch request hit at the Bloom filter: +4
score decrement on a prefetch request miss at the Bloom filter: -1
MIN-SCORE: minimum required score for selection of SPP, IP-stride, TSKID or MLOP: 0
NEXT-LINE-MIN-SCORE: minimum required score for selection of NL: 1500
EVAL-CNT: No of Prefetch activation events after which the selection process is repeated: 512
TSKID_SELECTION_ATTEMPT: Prefetch attempts made by TSKID which lead to its selection: 10000
BF-FPP: Required false positive rate of Bloom filter: 0.01
BF-EST-CAP: No. of entries in the Bloom filter: 2000

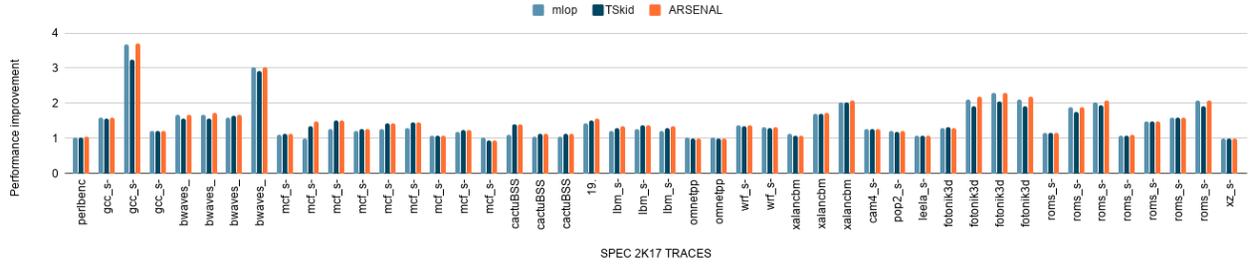

Figure 4: Normalized performance provided by Arsenal framework compared to components in test case 1.

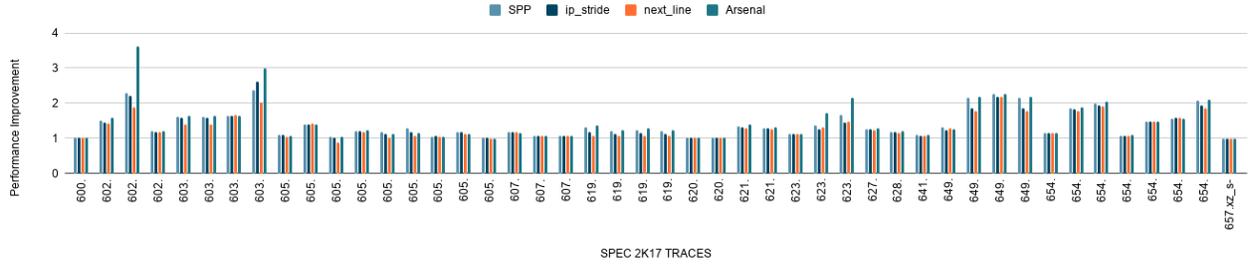

Figure 5: Normalized performance provided by Arsenal framework compared to components in test case 2

|  | entry-size × entries |  |
|---|---|---|
| Arsenal counters | Evaluate counters for N prefetchers (9xN bits) + Prefetch counters for N prefetchers (12xN bits) + prefetch scores for 3 prefetchers (11xN bits) | 0.03 KB |
| Bloom filters | 5 bits (false-positive-probability) + 32 bits (random-seed) + 11 bits (inserted-element-count) + 11 bit (projected-element-count) + 15 bit (table-size) + 3 bit (salt-count) + 2399*8 (bit table) + 135*32 (salt table) = 2948B × N | 2.87 KB |
| Thresholds |  | 0.1 KB |
| Total |  | 3 KB |

Table 1: Arsenal Hardware Overhead per component.

| Test case 1 |  |
|---|---|
| T-SKID overhead | 52.5 KB |
| MLOP overhead | 12 KB |
| Arsenal framework | 6 KB |
| Total | 70.5 KB |
| Test case 2 |  |
| SPP overhead | 5.73 KB |
| IP-stride overhead | 5.47 KB |
| Next line overhead | 0 KB |
| Arsenal framework | 9 KB |
| Total | 20.2 KB |

Table 2: Hardware Overhead for both test cases

| Mix no. | Mix details |
|---|---|
| 0 | 600.perlbench-570B 657.xz-2302B 605.mcf-994B 620.omnetpp-874B |
| 1 | 620.omnetpp-141B 641.leela-1083B 605.mcf-665B 607.cactuBSSN-4004B |
| 2 | 607.cactuBSSN-3477B 654.roms-1613B 623.xalancbmk-10B 605.mcf-1152B |
| 3 | 654.roms-1007B 628.pop2-17B 627.cam4-490B 605.mcf-782B |
| 4 | 607.cactuBSSN-2421B 605.mcf-1644B 619.lbm-4268B 619.lbm-2677B |
| 5 | 602.gcc-734B 605.mcf-1554B 619.lbm-3766B 605.mcf-472B |
| 6 | 649.fotonik3d-10881B 621.wrf-8065B 605.mcf-484B 619.lbm-2676B |
| 7 | 621.wrf-6673B 623.xalancbmk-165B 605.mcf-1536B 654.roms-293B |
| 8 | 654.roms-294B 602.gcc-1850B 603.bwaves-2931B 623.xalancbmk-202B |
| 9 | 649.fotonik3d-1176B 649.fotonik3d-8225B 654.roms-1070B 654.roms-523B |
| 10 | 654.roms-1390B 649.fotonik3d-7084B 603.bwaves-891B 602.gcc-2226B |
| 11 | 600.perlbench-570B 657.xz-2302B 605.mcf-1152B 654.roms-1007B |
| 12 | 628.pop2-17B 627.cam4-490B 619.lbm-2677B 602.gcc-734B |
| 13 | 605.mcf-472B 649.fotonik3d-10881B 619.lbm-2676B 621.wrf-6673B |
| 14 | 621.wrf-8065B 605.mcf-484B 623.xalancbmk-165B 605.mcf-1536B |
| 15 | 654.roms-294B 602.gcc-1850B 603.bwaves-1740B 603.bwaves-2609B |
| 16 | 654.roms-1070B 654.roms-523B 603.bwaves-891B 602.gcc-2226B |
| 17 | 600.perlbench-570B 657.xz-2302B 603.bwaves-891B 602.gcc-2226B |
| 18 | 654.roms-1070B 654.roms-523B 605.mcf-1644B 619.lbm-4268B |
| 19 | 603.bwaves-2609B 649.fotonik3d-1176B 607.cactuBSSN-2421B 605.mcf-1644B |
| 20 | 654.roms-1007B 619.lbm-2676B 603.bwaves-1740B 602.gcc-2226B |
| 21 | 605.mcf-1536B 605.mcf-1554B 605.mcf-1644B 605.mcf-994B |
| 22 | 603.bwaves-1740B 603.bwaves-2609B 603.bwaves-2931B 603.bwaves-891B |
| 23 | 649.fotonik3d-10881B 649.fotonik3d-1176B 649.fotonik3d-7084B 649.fotonik3d-8225B |
| 24 | 619.lbm-2676B 619.lbm-2677B 619.lbm-3766B 619.lbm-4268B |

Table 3: 25 representative 4-core mixes.

**Hardware Overhead:** Table 1 shows the hardware overhead of Arsenal framework. This is the overhead of the framework alone i.e. in addition to the memory overhead requirements of the component prefetchers.

Table 2 shows the hardware overhead for the two test cases taking into account the memory overhead requirements of the component prefetchers.

## 3. EVALUATION AND RESULTS

We used traces of SPEC CPU 2017 to evaluate the performance of Arsenal framework. Figure 4 shows the normalized performance with the Arsenal framework for test case1. On average, Arsenal provides 44.29% performance improvement, whereas MLOP and TSKID provide 38% and 40%, respectively. If we select the best prefetcher for each trace independently, then 43.76% improvement in performance is achieved. Thus this configuration outperforms IPCP [22], the winner of Data prefetching championship 3.

Figure 5 shows the normalized performance with the Arsenal framework for test case 2. On average, Arsenal provides 39.42% performance improvement, whereas SPP, next-line with degree 5, and IP-stride with degree 8 provide 35.1%, 32.6 and 31.89%, respectively. If we select the best prefetcher for each trace independently, then 35.92% improvement in performance is achieved.

The fact that the Arsenal framework outperforms even the ideal cases where we pick the maximum among speedups provided by its component prefetchers shows the effectiveness of our framework.

For multi-core evaluation, we created 25 representative mixes, as mentioned in Table 3. For these mixes, Arsenal provides an average speedup of 19.51% in test case 1 and 16.39% in case of test case 2.

## 4. CONCLUSION AND FUTURE WORK

This paper proposed the Arsenal framework that selects the best prefetcher from three prefetchers using a sandbox method. The framework uses Bloom filters to test the effectiveness of all the prefetchers. Arsenal provides an average performance improvement of 44.29% for the single-core traces. The effectiveness of Arsenal will improve if the framework gets multiple prefetchers that compliment each other: like a combination of regular and irregular prefetchers. Exploring the same along with modeling of DRAM contention for multi-cores is an exciting avenue for future work. Further research is also required to make the selection process adaptive so the framework can modify its selection criterion on the run if it encounters new workloads.

## 5. ACKNOWLEDGEMENT

Thanks to Biswabandan Panda, IIT Kanpur for his valuable suggestions.